\documentstyle[12pt]{article}
\newcommand{\half}{{1\over2}}
\newcommand{\be}{\begin{equation}}
\newcommand{\ee}{\end{equation}}
\newcommand{\bea}{\begin{eqnarray}}
\newcommand{\eea}{\end{eqnarray}}

\begin{document}
%%%%%%%%%%%%%Title page%%%%%%%%%%%%%%%%%%%

\begin{center}
\begin{large}
{\bf  Brick Walls on the Brane \\}
\end{large}  
\end{center}
\vspace*{0.50cm}
\begin{center}
{\sl by\\}
\vspace*{1.00cm}
{\bf A.J.M. Medved\\}    
\vspace*{1.00cm}
{\sl
Department of Physics and Theoretical Physics Institute\\
University of Alberta\\
Edmonton, Canada T6G-2J1\\
{[e-mail: amedved@phys.ualberta.ca]}}\\
\end{center}
\bigskip\noindent
\begin{center}
\begin{large}
{\bf
ABSTRACT
}
\end{large}
\end{center}
\vspace*{0.50cm}
\par
\noindent
The so-called ``brick-wall model'' is a semi-classical approach
that has been used to explain black hole entropy in terms of thermal
matter fields. Here, we apply the brick-wall formalism to thermal
bulk fields in a Randall-Sundrum brane world scenario. 
In this case, the black hole entity is really a string-like 
object in the anti-de Sitter bulk, while appearing
 as a Schwarzchild  black hole to observers
living on  the brane. In spite of these exotic circumstances,
we establish that the Bekenstein-Hawking entropy law is preserved.
Although a similar calculation was recently considered in the
literature, this prior work invoked a simplifying assumption
(which we avoid) that can not be adequately justified.
%PACS 04.70.Dy
\newpage

\section{Introduction}\medskip
\par
Much literary attention has recently been directed to the notion
that our ``physical'' universe is really just a 3+1-dimensional
submanifold (i.e., three brane) which is embedded in a 4+$n$-dimensional
bulk \cite{add}. Particularly interesting proposals  along this line
have originated from the work of Randall and Sundrum \cite{rs1,rs2}.
These authors considered a 5-dimensional anti-de Sitter (AdS) spacetime,
with the  ``extra'' bulk dimension being related to the
3+1-submanifolds via a ``warped'' compactification factor.
The first of these models, RS1 \cite{rs1}, utilizes a pair of branes (one with
positive tension and the other,  negative)
such that the physical universe is embedded on the negative-tension brane.
In their second proposal, RS2 \cite{rs2}, the universe lives on a single
 brane of positive tension.\footnote{Alternatively, one can regard
RS2 as a dual-brane system in which the negative-tension brane
has been moved out to the AdS horizon.}
The phenomenological and cosmological implications of such brane-world 
 scenarios have been considered in a multitude of studies. 
(See Ref.\cite{rub} for a review and references.)
\par
One of the more interesting aspects of the RS brane world is how it 
may influence the physics of black holes.
It is clear that the gravitational collapse of matter will
result in the formation of a black hole (or, at least, 
a black hole like object)
 which,  from a brane  perspective,  must maintain
its usual astro-physical properties.
However, from a bulk perspective, this black entity must be
viewed as a 5-dimensional extended object, as gravitons (unlike
most particles) are free to propagate through the extra 
dimensions of spacetime.  In an attempt to resolve these
paradoxical implications, CHR (Chamblin, Hawking and Reall \cite{chr}) have
proposed a 5-dimensional black string, whereby   the  induced
metric on the brane  reduces to   the standard  Schwarzchild
solution. 
\par
As CHR have  pointed out themselves, 
the black string solution suffers from instabilities
near the AdS horizon \cite{chr}. However, they overcame this
obstacle by virtue of the following argument. 
It is known that the black string  can also be
unstable due to  perturbations of  wavelength on the order
of the horizon radius, $r_{h}$. (This effect is commonly known as
the Gregory-Laflamme instability \cite{gl}.) However, the AdS
bulk geometry can act as a confining box that prevents
fluctuations greater than $l$ from developing
(where $l$ is the usual AdS length parameter). That is, stability will
be maintained as long as $l<r_{h}$.  Now consider that, as an artifact
of the RS geometry, the black hole effective mass  ($M_{e}$)
  decreases exponentially
with transverse proper distance ($y$) away from the brane. Hence,
$r_{h} \sim  M_{e}$  must decrease below $l$ at some point
along the $y$-axis.  As conjectured by CHR, at  this point, 
 the black string will  ``pinch off'' and form into a stabilized ``black
cigar''. Simple arguments have since  verified that the transverse
extent of the cigar will be small enough to  avoid 
the unfavorable complications of the AdS horizon \cite{gkr,cha}.
\par
Much work has already been done in generalizing the CHS solution,
as well as examining some of its thermodynamic properties.
(For an extensive but incomplete list, see Ref.s\cite{gkr}-\cite{kor}.)
The purpose of our current paper is to further this intriguing topic by
way of 't Hooft's so-called ``brick-wall model''  \cite{th}.
\par
The underlying premise of the 't Hooft  methodology \cite{th} is
that  the Bekenstein-Hawking black hole entropy 
($S_{BH}=\pi r_{h}^2/l_{p}^{2}$ where $l_{p}$ is the Planck length 
\cite{bek,haw}) can be accounted for via the statistical entropy
of thermal fields (particularly, those near the event horizon). 
For such an approach to
be viable, it is necessary to introduce an artificial boundary,
or ``brick wall'', just outside of the black hole horizon.
This wall controls the ultraviolet divergences that are inherent
to  this type of calculation. Although a seemingly unphysical procedure,
the introduction of a wall  can be justified as follows: quantum fluctuations
prevent events within a Planck length of the horizon from being seen
by an external observer.
\par
Since the inception of the brick-wall model, many authors have
applied it to various black hole geometries. (For yet another
extensive, incomplete list, see Ref.s\cite{BW1}-\cite{ew}.)
This model  has also endured much constructive criticism;
see Ref.\cite{mi} for an interesting discussion and references.
A pair of relatively  recent papers, however, have significantly improved
the status of brick-wall calculations.
\par
Firstly, Mukohyama and Israel \cite{mi} have resolved many of the
critical issues; including massive energy densities arising near
the horizon, the unphysical implications of an artificial wall,
and the back reaction of this effective boundary on the black hole
geometry. They accomplished this by identifying the ground state
of the brick-wall model as a ``topped-up'' Boulware state \cite{bw}
(i.e., the Boulware vacuum plus thermal excitations). It can
be consequently argued that the presence of a brick wall with
thermal excitations is an alternative, equivalent description
of the Hartle-Hawking vacuum state \cite{hh}. That is, the entropy
of the thermal fields just outside of the wall can be
identified with the geometrical entropy that arises out of
the Gibbons-Hawking ``instanton'' \cite{gh}.
\par
Secondly, Winstanley \cite{ew} has rigorously demonstrated
that, at least for large black holes, the brick-wall entropy
can be entirely accounted for by renormalizing the coupling constants
of the ``complete''
one-loop effective action.\footnote{This gravitational  action is complete in
the sense that it includes terms that are quadratic in the curvature.
We also note that such a program of renormalization was
originally proposed in Ref.\cite{ren,woops1} and had already
been demonstrated via a conical-singularity method of renormalization
\cite{woops2}.}
Not only was this demonstrated for the divergent  entropy terms
(i.e., terms that diverge as the brick wall coincides with the horizon),
but for the finite terms as well.  Although the procedure broke down
with regard to small black holes, this failure can be attributed to 
quantum gravity corrections, which become important in this limiting case.
\par
This  brief discussion on branes and brick walls leads us 
to the focus of the current study. Namely, the contribution of
thermal bulk fields to the entropy of a  black
hole on a Randall-Sundrum brane or, alternatively, a  black cigar in the 
bulk. (Note that the contribution
of thermal brane fields would proceed as in any number of
prior publications, starting with Ref.\cite{th}.) 
Our particular interest is to see if the Bekenstein-Hawking area
law \cite{bek,haw} 
is preserved in the leading-order divergent term(s). A failure
in this regard could jeopardize the renormalization process as
discussed directly above.
\par
 The rest
of this paper thus proceeds as follows. Section 2  considers
the necessary formalism for the calculations of interest. 
In Section 3, we evaluate    the free energy associated with a thermal
bulk field. As shown in Section 4,  it is then   straightforward to extract
the corresponding  entropy.  In Section 5, we consider
the thermal energy and use this result to touch base with
an earlier study on thermal fields in the Randall-Sundrum brane
world \cite{xxx}.
The paper ends in Section 6 with a summary and discussion of the results.
\par
Before concluding this introductory section, we point
out that a similar study (to ours) has been carried out 
 by Kim et al. \cite{kor}.
However, their analysis relied on a unreasonable assumption 
in order to simplify the calculations. This point will be elaborated
on at an appropriate interval in our paper.

\section{The Setup}

We begin here by considering a dual-brane Randall-Sundrum 
scenario in 5-dimensional
AdS. Without loss of generality, we place 
the positive-tension brane at $y=0$ (with $y$ denoting the extra
bulk dimension).  
 This would be just like RS1 \cite{rs1}, except that we 
assume the physical universe
to be living on the positive-tension brane. Hence, the model of interest
is more in the ``spirit'' of RS2 \cite{rs2}; however, in this study, it is
necessary to cut off the bulk spacetime at some point $y_{c}$ as will
be explained below. 
\par
If we further assume Poincare invariance on the branes, then the
general solution can be written as follows \cite{rs1,rs2}:
\be 
ds^2=e^{-2ky}\left[g_{\mu\nu}dx^{\mu}dx^{\nu}\right]+dy^2,
\label{0}
\ee
where the inverse AdS length parameter, $k$, has been appropriately
fixed in terms of the cosmological constant and brane tensions,
and where $g_{\mu\nu}$ describes a 3+1-dimensional  Ricci-flat spacetime.
\par
It is a common practice to take $g_{\mu\nu}$ to be the 3+1-Minkowski
metric. However, since our interest is in black holes, we
follow CHR \cite{chr} and incorporate a 3+1-Schwarzchild geometry.
That is: 
\be
ds^2=e^{-2ky}\left[-U(r)dt^2+U^{-1}(r)dr^2+r^2d\theta^2+r^2\sin^2\theta
d\phi^2\right]+dy^2,
\label{1}
\ee
where $U(r)=1-2MG^{(4)}/r=1-r_{h}/r$. This solution describes a
black string in 5-dimensional AdS.
\par
As discussed in Section 1 (and see Ref.\cite{chr}), 
the black string is expected
to pinch off (on account of the Gregory-Laflamme instability \cite{gl})
 and form into a stable ``black cigar''
well before reaching the AdS horizon. That is, Eq.(\ref{1}) can be
considered an approximate solution with validity over some
finite region $|y|<y_{p}$, where $y_{p}$ represents the ``pinching-off''
 point.\footnote{A precise evaluation of $y_{p}$ remains an unresolved
problem. However, ``ballpark'' estimates have put it at $y_{p}\sim
k^{-1}\ln(kr_{h})$ \cite{gkr,cha}.} Hence, we will effectively restrict
the bulk spacetime by placing a second brane at $y=y_{c}$, where
$y_{c}<y_{p}$ is to be assumed. Let us further assume $Z_{2}$ symmetry,
and so considerations may be limited to the region $0\leq y\leq y_{c}$.
\par
Since we are following the brick-wall program of 't Hooft \cite{th}, it
is appropriate to consider a matter field propagating in the
relevant spacetime. For  simplicity, let us assume a minimally coupled
(massive) scalar field that satisfies the Klein-Gordon equation:
\be
\Box^{(5)} \Psi -m^2\Psi=0.
\label{2}
\ee
Rewriting this expression in terms of Eq.(\ref{1}), we have:
\bea
{e^{2ky}\over r^2} \partial_{r}\left[r^2 U(r)\partial_{r}\Psi\right]
- {e^{2ky}\over U(r)} \partial^{2}_{t} \Psi + {e^{2ky}\over r^2 \sin\theta}
\partial_{\theta}\left[\sin\theta\partial_{\theta}\Psi\right]
+ {e^{2ky}\over r^2 \sin^{2}\theta} \partial^{2}_{\phi}\Psi
\nonumber \\ 
+{1\over e^{-4ky}}\partial_{y}\left[e^{-4ky}\partial_{y}\Psi\right]
-m^2\Psi=0.
\label{3}
\eea
\par
In compliance with the 't Hooft prescription, it is also
necessary to introduce the following boundary conditions:
\be
\Psi=0 \quad\quad for \quad\quad r\leq r_{h}+\epsilon ,
\ee
\be
\Psi=0 \quad\quad for \quad\quad r\geq L.
\ee
Here,  $\epsilon<<r_h$ represents the brick-wall cutoff 
 that eliminates the ultraviolet divergences;
whereas the boundary at $L>>r_{h}$ eliminates the infrared divergences.
\par
Given the spherical symmetry of the 4-dimensional brane world and
the existence of a timelike Killing vector,
the scalar field can be decomposed as follows:
\be
\Psi=e^{-iEt}Y_{l,m_l}(\theta,\phi)f(y)R(r).
\ee
$Y_{l,m_l}(\theta\phi)$ is the usual spherical harmonic function,
which is known to satisfy:
\be
{1\over\sin\theta} \partial_{\theta}\left[\sin\theta\partial_{\theta}
Y_{l,m_l}(\theta,\phi)\right]+{1\over\sin^{2}\theta}\partial^{2}_{\phi}
Y_{l,m_l}(\theta,\phi)=-l(l+1)Y_{l,m_l}(\theta,\phi).
\ee
\par
Let us now define an ``effective mass'' $m_{n}$, where $n$ labels the
various modes   of the function $f(y)$, such that:
\be
{1\over e^{-4ky}}\partial_{y}\left[e^{-4ky}\partial_{y}f_{n}(y)\right]
-m^2f_{n}(y)=-e^{2ky}m_{n}^2f_{n}(y).
\label{8}
\ee
Then Eq.(\ref{3}) conveniently reduces to the following radial equation:
\be
{1\over r^2} \partial_{r} \left[r^2 U(r) \partial_{r}R(r)\right]
+{E^2\over U(r)} R(r)-{l(l+1)\over r^2}R(r)-m_{n}^2 R(r)=0.
\label{9}
\ee
\par
To further simplify the separated wave equations (\ref{8},\ref{9}),
 we can invoke the WKB approximation. That is, we now assume
that each of  $R(r)$ and $f_{n}(y)$ can be expressed as the product
of a slowly varying amplitude and an exponent with a rapidly
varying phase. To leading order, one need only consider the derivatives of the
phase functions, and this leads to the following expressions:
\be
-{1\over r^2 U(r)}\partial_{r}\left[r^2 U(r)\partial_{r}R(r)\right]=
K^2_{r}R(r), 
\label{10}
\ee
\be
-{1\over e^{-4ky}}\partial_{y}\left[e^{-4ky}\partial_{y}f_{n}(y)\right]=
K^2_{n}f_{n}(y); 
\label{11}
\ee
where the wave numbers (each corresponding to the derivative of
the appropriate phase) are given as follows:
\be
K_{r}={1\over U(r)}\left[E^2 -U(r)\left({l^2+l\over r^2}+m_n^2\right)
\right]^{\half},
\label{12}
\ee
\be
K_{n}=\left[m_n^2e^{2ky}-m^2\right]^{\half}.
\label{13}
\ee
\par
What will be particularly useful is the degeneracy  of modes ($n_i$) for
any given wave number ($K_i$). According to the semi-classical quantization
rule, we have \cite{th}:
\be
n_{r}={1\over\pi}\int^{L}_{r_h +\epsilon}drK_{r},
\label{14}
\ee
\be
n_{n}={1\over\pi}\int^{y_c}_{0}dyK_{n}.
\label{15}
\ee
From Eqs.(\ref{13},\ref{15}), it is straightforward to obtain the following
useful result:
\be
{dn_{n}\over dm_{n}}={1\over\pi k m_n}\left[\sqrt{m_n^2 e^{2ky_c}-
m^2}-\sqrt{m_n^2-m^2}\right].
\label{16}
\ee

\section{The Free Energy}

Let us now proceed to evaluate the free energy ($F$) of
 a thermal bath of bulk scalars at temperature $\beta^{-1}$.
We begin by considering the standard thermodynamic definition:
\be
e^{-\beta F}= \sum_{\tau}e^{-\beta E_{\tau}},
\label{17}
\ee
where $E_{\tau}$ is the thermal energy corresponding to
quantum state $\tau$. Since the analysis is for bosons (whose
occupation number can take on any positive integral value or
zero), the following is an equivalent relation:
\be
e^{-\beta F}=\prod_{n_{r}, n_{n}, l, m_l} \left(1-e^{-\beta E}\right)^{-1}.
\label{18}
\ee
\par
Solving for $F$, we have:
\be
F={1\over\beta}\sum_{n_r,n_n,l,m_l} \ln\left(1-e^{-\beta E}\right)
\label{19}
\ee
or in the continuum limit:
\be
F={1\over\beta}\int dl(2l+1)\int dn_n \int dn_r \ln\left(1-e^{-\beta E}
\right).
\label{20}
\ee
The factor of $2l+1$ is, of course, due to the degeneracy of
the quantum number $m_l$
for a given value of $l$.
\par
After some additional manipulation, including an integration  by parts,
we find:
\be
F=-\int dl(2l+1)\int dm_n \left({dn_n\over dm_n}\right)\int dE {1\over
e^{\beta E}-1} n_r.
\label{21}
\ee
The substitution of  Eqs.(\ref{12},\ref{14},\ref{16}) then yields:
\bea
F=-{1\over \pi^2 k}\int dl (2l+1)\int dm_n {1\over m_n} 
\left[\sqrt{m_n^2 e^{2ky_c}-
m^2}-\sqrt{m_n^2-m^2}\right] \nonumber \\
\times  \int dE {1\over e^{\beta E}-1} \int
^{L}_{r_h + \epsilon} dr {1\over U(r)} 
\sqrt{E^2 -U(r)\left({l^2+l\over r^2}+m_n^2\right)}.
\label{22}
\eea
\par
Note the unspecified limits of integration
in the above expression. We must  integrate   over all values
 of phase space for which the reality of the square roots is preserved.
For the implied order of integration, this condition leads to
the following limits:
\be
0\leq l\leq \half \left[-1+\sqrt{1+4r^2\left({1\over U(r)}E^2-m_n^2
\right)}\right],
\label{23}
\ee
\be
m\leq m_n\leq{E \over \sqrt{U(r)}},
\label{24}
\ee
\be
m\sqrt{U(r)}\leq E\leq\infty.
\label{25}
\ee
\par
Before proceeding any further, let us consider  the integration over
all permissible values of the mass parameter $m_n$. This is
the primary difference between our calculation and that of a
prior study \cite{kor}. The authors of Ref.\cite{kor}
fixed the effective 4-dimensional mass $m_n$ (which they called
$\mu$) equal to
the 5-dimensional mass $m$.\footnote{More accurately, they fixed $m_n$
(or $\mu$) equal to $me^{-ky}$ and then used $y=0$ on the brane.}
Such a simplification is contrary to the results of
prior works that have studied the decomposition of bulk scalar
fields. (See, for instance, Ref.\cite{gw1}.) It has been amply
demonstrated that the bulk field manifests itself, to
a 4-dimensional observer, as an infinite tower of scalars;
each of which has an associated, distinct 
value of mass. (Note that these masses
are obtainable, in principle, by solving the relevant eigenvalue
problem.) Hence, for a reliable calculation, there can be no justification 
in singling
out any one particular value of $m_n$ as the preferred one.
\par
The $l$ integration of Eq.(\ref{22}) can be done explicitly to yield:
\bea
F=-{2\over 3\pi^2 k} \int^{E/\sqrt{U(r)}}_{m} dm_n {1\over m_n} 
\left[\sqrt{m_n^2 e^{2ky_c}-
m^2}-\sqrt{m_n^2-m^2}\right]  \nonumber \\
\times
 \int^{\infty}_{m\sqrt{U(r)}} dE {1\over e^{\beta E}-1} \int
^{L}_{r_h + \epsilon} dr \left[E^2-U(r)m_n^2\right]^{3\over 2}.
\label{26}
\eea
\par
Let us now consider the integration over $m_n$.  This 
can not be done exactly, except in the trivial case $m=0$. However,
we can obtain a very reasonable approximation by first considering
the following argument. 
\par
Ultimately, we are only interested
in contributions to the entropy (and, hence, free energy)
arising from the proximity of the wall. 
The remaining contribution, for which $r>>r_{h}$,
is well understood to be the entropy of a quantum field
in flat space \cite{th} and can be neglected 
for our purposes. With this, as well as $U(r)\approx 0$ if $r\approx r_{h}$, 
 in mind, the following are
valid approximations for quantities in the above integrand: 
\be
E^2-U(r)m_n^2\approx E^2 \quad\quad unless \quad\quad 
m_{n}\approx E/\sqrt{U(r)},
\label{27}
\ee
\bea
{1\over m_n} 
\left[\sqrt{m_n^2 e^{2ky_c}-
m^2}-\sqrt{m_n^2-m^2}\right] \approx e^{ky_c}-1 
\nonumber \\
 \quad\quad\quad\quad unless 
\quad\quad 
m_n\approx m.
\label{28}
\eea
\par
By virtue of the above observations, it is not difficult to show that:
\bea
\int^{E/\sqrt{U(r)}}_{m} dm_n {1\over m_n} 
\left[\sqrt{m_n^2 e^{2ky_c}-
m^2}-\sqrt{m_n^2-m^2}\right] 
 \left[E^2-U(r)m_n^2\right]^{3\over 2}
\nonumber \\
\approx 
m \Upsilon(ky_c)E^3 +{3\pi\over 16}{(e^{ky_c}-1)\over \sqrt{U(r)}} E^4,
\label{30}
\eea
where:
\be
\Upsilon(ky_c)\equiv 2\tan^{-1}\left(e^{ky_c}+\sqrt{e^{2ky_c}-1}
\right)-{\pi\over 2} -\sqrt{e^{2ky_c}-1}. 
\label{31}
\ee 
Again, we note that Eq.(\ref{30}) is exact when $m=0$.
\par
Substituting Eq.(\ref{30}) into Eq.(\ref{26}), we can see that
the remaining integrations, over $E$ and $r$, have been separated.
To perform the integration over $E$, it is useful to note that
the lower bound tends to zero when near the horizon; cf. Eq.(\ref{25}). 
Hence,  standard formulas \cite{for} can be applied to obtain the
following:
\be
\int^{\infty}_{0} dE{E^4\over e^{\beta E}-1}= {24\zeta(5)\over\beta^5},
\label{32}
\ee
\be
\int^{\infty}_{0} dE{E^3\over e^{\beta E}-1}= {\pi^4\over 15\beta^4}.
\label{33}
\ee
\par
With the above set of results, Eq.(\ref{26}) simplifies  as follows:
\be
F_{\epsilon}\approx -{3\zeta(5)(e^{ky_c}-1)\over
\pi k \beta^5}\int_{r_h +\epsilon} dr {r^2\over \left[U(r)\right]^{5/ 2}}
-
{\pi^2 m \Upsilon(ky_c)\over
45 k \beta^4}\int_{r_h +\epsilon} dr {r^2\over \left[U(r)\right]^2},
\label{34}
\ee
where the subscript $\epsilon$ indicates a near-horizon form.
\par
Because of our near-horizon considerations,  
 the following approximation can be used:
\be
U(r)\approx (r-r_h)\left.{dU\over dr}\right|_{r=r_h}={(r-r_h)\over r_h}.
\label{34A}
\ee
The integration of Eq.(\ref{34}) is now straightforward and yields:
\be
F_{\epsilon}\approx - {2\zeta(5)(e^{ky_c}-1)\over \pi
k\beta^5} {r_h^{9/ 2}\over \epsilon^{3/2}}
-
{m\pi^2\Upsilon(ky_c)\over 45 
k\beta^4} {r_h^{4}\over \epsilon}.
\label{35}
\ee

\section{The Entropy}

We finish off the analysis by calculating the near-horizon
contribution to  the entropy. The first law of thermodynamics tells
us:
\be
S_{\epsilon}=\beta^2 {\partial F_{\epsilon} \over \partial \beta}.
\label{35A}
\ee
\par
From a thermodynamic perspective, it is most appropriate to
make an off-shell evaluation and then consider the on-shell
limit \cite{BW4}. Hence, we directly apply Eq.(\ref{35A})  to 
Eq.(\ref{35}) to obtain:
\be
S_{\epsilon}\approx  {10\zeta(5)(e^{ky_c}-1)\over \pi
k\beta^4} {r_h^{9/ 2}\over \epsilon^{3/2}}
+
{4 m\pi^2\Upsilon(ky_c)\over 45 
k\beta^3} {r_h^{4}\over \epsilon}
\label{36}
\ee
and then use the well-known on-shell Schwarzchild 
relation\footnote{The Schwarzchild relation is appropriate given that
the surface gravity (and, hence, temperature) is
constant   at all points along the event horizon; whether on the
brane or in the bulk \cite{BH1}.}
of 
$\beta=4\pi r_h$ \cite{gh}. This yields:
\be
S_{\epsilon}\approx  {5\zeta(5)(e^{ky_c}-1)\over 128\pi^5
k} {r_h^{1/ 2}\over \epsilon^{3/2}}
+
{ m\Upsilon(ky_c)\over 720 \pi 
k} {r_h\over \epsilon}.
\label{37}
\ee
\par
To make sense of this entropy result, it is necessary to re-express
$\epsilon$ in terms of the invariant distance from  the horizon
to the brick wall \cite{th}. Denoting this invariant distance as
${\tilde \epsilon}$, we can write:
\be
{\tilde \epsilon}=\int^{r_h +\epsilon}_{r_h} dr {1\over \sqrt{U(r)}}.
\label{38}
\ee
With the help of Eq.(\ref{34A}), this becomes:
\be
{\tilde \epsilon}^2=4 r_h \epsilon .
\label{39}
\ee
So, in terms of invariant quantities only, the  near-horizon
entropy (\ref{37}) takes on the form:
\be
S_{\epsilon}\approx \left[ {5\zeta(5)(e^{ky_c}-1)\over 16\pi^6
k} {1\over {\tilde\epsilon}^3}
+
{ m\Upsilon(ky_c)\over 180 \pi^2 
k} {1\over {\tilde\epsilon}^2}\right] {A_h\over 4},
\label{40}
\ee
where $A_h=4\pi r_h^2$ is the horizon area on the brane. 
Notice that  the square-bracket quantity appears to be
independent of the black hole geometry\footnote{A possible
exception to this statement may be  
the parameter $y_c$, depending on its physical interpretation.
This point is elaborated on in Section 5.}. Thus,
we have successfully verified  the black hole 
area law with regard to  contributions from  bulk scalar fields.
We further interpret this result in the  Section 6.

\section{The Thermal Energy}

In this brief section, we compare our results with an earlier
study by  Brevik et al. \cite{xxx}. These authors  considered
thermal quantum fields in a ``conventional'' (i.e., non-black hole)
Randall-Sundrum  setting. In this prior work, the thermal
energy was evaluated for both high and low temperature 
limits. Naturally, the low-temperature regime is most suitable
for comparison with a semi-classical black hole. We thus quote
their thermal-energy result (as applicable to scalar fields) 
for a large value of inverse temperature $\beta$ \cite{xxx}:
\be
{\cal E}=\sum_{n}{V_3m_n^{{5\over 2}}e^{-\beta m_n}\over 
(2\pi\beta)^{3/ 2}},
\label{111}
\ee
where the summation is over all possible ``Kaluza-Klein-like'' 
modes\footnote{That is, modes that arise in the decomposition
of the bulk scalar field. Hence, $m_n$ is the ``effective mass''
as defined by  Eq.(\ref{8}).} and $V_3$ is
the 3-dimensional brane volume of interest.
\par
Because of the rapidly vanishing exponential factor, 
it is a suitable approximation
to set $m_n$ equal to its lower-bound value. (This is what was
effectively done in Ref.\cite{xxx}.) Hence, by way of
Eq.(\ref{24}), the above can be simplified as follows:
\be
{\cal E}\approx {V_3 m^{{5\over 2}}e^{-\beta m}\over 
(2\pi\beta)^{3/ 2}}.
\label{222}
\ee
\par
Let us now consider the thermal energy in our black hole model.
This is directly obtainable from the free energy (\ref{35})
with application of a standard thermodynamic relation:
 ${\cal E}=\partial(\beta F)/\partial\beta$. This  yields:
\be
{\cal E}_{\epsilon}\approx  {8\zeta(5)(e^{ky_c}-1)\over \pi
k\beta^5} {r_h^{9/ 2}\over \epsilon^{3/2}}
+
{m\pi^2\Upsilon(ky_c)\over 15 
k\beta^4} {r_h^{4}\over \epsilon}.
\label{333}
\ee 
Next using ${\tilde \epsilon}^2=4r_h\epsilon$, $\beta=4\pi r_h$
and appropriately setting $V_{3}=4\pi r^2_h {\tilde\epsilon}$,
we find:
\be
{\cal E}_{\epsilon}\approx {V_3\over\beta}
\left[ {\zeta(5)(e^{ky_c}-1)\over 16 \pi^6
k  {\tilde\epsilon}^4}
+
{m\Upsilon(ky_c)\over 960 \pi^2
k {\tilde\epsilon}^3}\right].
\label{444}
\ee 
\par
A comparison of Eq.(\ref{444}) with Eq.(\ref{222}) reveals quite
a contrast in thermodynamics for the two different
scenarios. For instance, only the black hole thermal
energy has an explicit dependence on the bulk parameter
$y_c$  in this low-temperature regime. (Although, one would expect
this to change when considering generic values of temperature.)
Also, it is evident that the black hole thermal energy
diverges much more slowly as $\beta\rightarrow \infty$. 
This behavior  can likely be attributed to the black hole
horizon being a surface of infinite ``red-shifting''.
That is, only modes of arbitrarily small wavelength can exist
close to the wall \cite{th,mi}.

\section{Conclusion}

In the preceding paper, we have considered a bulk scalar field
propagating in the background spacetime of a black hole on
a  Randall-Sundrum
brane \cite{rs1,rs2} (or black cigar in the bulk \cite{chr}). 
A semi-classical quantization procedure, along the lines
of 't Hooft's original brick-wall model \cite{th}, allowed
us to calculate the thermal energy due to this field. The
near-horizon contribution to this thermal energy  lead
directly to an evaluation of
 the corresponding entropy. This result (see Eq.(\ref{40})) was found to
satisfy the Bekenstein-Hawking area law of black hole entropy \cite{bek,haw}.
We also considered the thermal energy, and how it compared
with that found in an earlier study \cite{xxx}.
\par
The complete black hole entropy in this RS brane world is
 obtainable, in principle, by summing over the thermal contributions of
all relevant fields. This summation should include bulk fields, as well as
those restricted to the brane (i.e., the Standard Model fields).
The leading-order divergent term (or terms) could then be 
absorbed in a renormalization of Newton's gravitational constant
to yield the standard form of $A_{h}/ 4 G^{(4)}$ \cite{ren}.
Meanwhile, the sub-leading divergent terms could (in principle) be absorbed by
renormalizing  the coupling constants of action terms that are
quadratic in  curvature \cite{ew}. 
\par
With the above discussion in mind, it is interesting to
compare the leading-order contribution from the bulk fields 
($S_{bu}\sim A_h {\tilde\epsilon}^{-3}$) with the analogous result
for  brane fields of $S_{br}\sim A_{h} {\tilde\epsilon}^{-2}$
\cite{th}.
This comparison would imply that the bulk fields dominate
the entropy, but this need not be the case. To see this, first consider that
 the bulk-field
 contribution also contains a factor of $k$ in the denominator 
(i.e., the inverse AdS parameter). Because of hierarchical
arguments, it is believed that
$k^{-1}$ is on the order of the Planck length \cite{rs1,rs2}.
One would also expect the ultraviolet cutoff ${\tilde\epsilon}$
to be of this order, as quantum fluctuations prevent
events closer (to the horizon) than $l_{pl}$  
from being observed \cite{mi}. Hence, the leading-order contributions 
from  bulk and brane fields should be of the same order;  namely,
$S_i\sim r_h^2 l_{pl}^{-2}$.
\par
It is interesting to note that the sub-leading term
in the bulk thermal entropy is likely negative, 
as can been seen by carefully examining the defining relation
for $\Upsilon(ky_c)$; cf. Eqs.(\ref{31},\ref{40}).
If $m\sim m_{pl}$, then this term is also of order $r_{h}^2 l_{pl}^{-2}$
and may effectively cancel out the  ${\tilde\epsilon}^{-3}$ contribution.
It is worth pointing out, however, that the bulk fields (for instance:
gravitons,
three-form tensors, moduli scalars) are  likely to be
predominantly massless; thus, negating this negative contribution.
\par
For illustrative purposes, we have focused attention on
 a relatively simple
brane world scenario. However, the brick-wall formalism should be
applicable to other, (perhaps) more interesting cases. For instance,
one might apply these techniques to a model where  the brane is
realized dynamically out of a  higher-derivative gravity bulk \cite{yyy,zzz}.
Significantly to this case, the bulk can     be  interpreted as a 5-dimensional
 Schwarzchild-anti-de Sitter black
hole.  In Ref.\cite{zzz}, Nojiri et al. calculated the associated 
 entropy (by geometric arguments) and found that, in general,
there was a discrepancy between this  entropy and that deduced
from holographic considerations in 4-dimensions.\footnote{This
``holographic'' entropy was
obtained \cite{zzz} by identifying the brane dynamics as a 
Friedmann-Robertson-Walker
cosmological equation and then  applying an analogue of the Cardy formula
\cite{www}.} The authors then  interpreted this discrepancy as a measure
of the deviation from a broken AdS/CFT correspondence \cite{ads}.
It would be interesting to see if this discrepancy is resilient in a
 brick-wall context, and we hope to address this
issue in a future work. We do, however, anticipate that
the discrepancy perseveres, given the observed breakdown in      
 the area law   for a 5-dimensional
brick-wall calculation  \cite{BW8}
(noting that a similar breakdown was found in Ref.\cite{zzz}).
\par
Finally,  a brief comment regarding the position of the negative-tension
brane (i.e., $y_c$) is in order. If we assume an explicit RS2 model (where 
the negative brane tends to the AdS horizon), then the parameter
$y_c$ (as it appears in our formalism) no longer represents
the location of the negative brane.  Rather, it  represents
a transverse limit in the bulk due to instabilities in the
black string solution  \cite{chr,gl}. If this is the case, $y_c$
would  not be determined by a stabilization mechanism (such as
those of Ref.\cite{gw2,gw3}); instead, it should probably 
be regarded as an implicit function of the black hole
geometry. So in this event, the area law breaks down,
as  both terms in Eq.(\ref{40}) depend explicitly on $y_c$.
However, one could argue that the multiplicity of bulk fields is
small compared to that of the brane fields (which includes all
particles  prescribed by the Standard Model); thus, suppressing
the bulk contribution to the entropy. That is to say, observers living
 on the brane
may not readily detect such a breakdown.

\section{Acknowledgments}
\par
The author  would like to thank  V.P.  Frolov  for helpful
conversations. 
  \par\vspace*{20pt}

%\newpage

\end{document}